\documentclass[12pt]{iopart}
\input{psfig.sty}
\usepackage{graphicx}
\begin{document}

\title [Collective oscillations of a quasi 1D BEC under damping]
{Collective oscillations of a quasi one dimensional Bose
condensate under damping}

\author{Fatkhulla Kh Abdullaev $^{1,}$ $^2$, Ravil M. Galimzyanov
$^2,$ \footnote[7] {To whom correspondence should be addressed}
and Khayotullo Ismatullaev $^{2,}$ $^3$ }

\address{$^{1 }$ Dipartamento di Fisica "E.R. Caianiello",
Universit\'a di Salerno, I-84081 Baronissi (SA), Italy}

\address{$^{2 }$ Physical-Technical Institute of the Academy of
Sciences, G.Mavlyanov 2-b, Tashkent 700 084, Uzbekistan}

\address{$^{3 }$ Institute of Electronics of the Academy of
Sciences, F.Khodjaev 33, Tashkent 700 125, Uzbekistan}
\ead{ravil@uzsci.net}

\begin{abstract}
Influence of the damping on collective oscillations of a
one-dimensional trapped Bose gas in the mean field regime has been
studied. Using the phenomenological damping approach developed by
L.P. Pitaevskii, modified variational equations for the parameters
of the condensate wave function is derived. Analytical expressions
for the condensate parameters in equilibrium state have been
obtained. Bistability in nonlinear oscillations of the condensate
under periodic variations of the trap potential is predicted. The
predictions of the modified variational approach are confirmed by
full numerical simulations of the 1D GP equation with the damping.
\end{abstract}


\submitto{jpb}

\maketitle

\section{Introduction}
The dynamics of a one-dimensional trapped ultra-cold Bose gas has
attracted considerable attention for last years \cite{PS}.
Recently has been achieved regimes where 1D Bose gas with a
zero-range two-body interaction in the Tonks -Girardeau (TG)regime
\cite{LL,L,G} with fermionic behavior becomes visible. An
investigation of the transition from the mean field regime to the
TG regime was performed in \cite{Menotti}. Experimentally 1D
regime has been realized in \cite{exp}.

Measurements  of the collective oscillations of such a system
should give a lot of information about the BEC dynamics. In
particular this is important in the analysis of the condensate
dynamics in a magnetic waveguide, being a fundamental atom optical
element \cite{Ott}. Performed by this time theoretical
descriptions have mainly dealt with conservative systems, e.g. see
\cite{AG}, where collective oscillations of a quasi 1D
Bose-Einstein condensate(BEC) in the low and high density regimes
were investigated. The damping of the radial BEC oscillations in a
cylindric trap, connected with the parametric resonance and
leading to the energy transfer from collective oscillations to
longitudinal sound waves has been studied in \cite{Kagan}. The
dissipation inheres in real systems. So it is of interest to
investigate theoretically effect of damping on collective
oscillations of a one-dimensional trapped repulsive Bose gas.

We consider here the problem using the phenomenological approach
developed by L.P. Pitaevskii \cite{LPP} and employed later in
\cite{Burnett}

\section{The model}
The dynamics of a trapped one dimensional repulsive Bose gas with
the damping is described in the framework of the 1D
Gross-Pitaevskii equation \cite{LPP,Burnett}

\begin{equation}\label{gpe}
i\hbar\phi_{t} = (1+i\gamma)(-\frac{\hbar^2}{2m}\phi_{xx}
+V(x,t)\phi + g_{1D}|\phi|^{2}\phi -\mu\phi),
\end{equation}
with the total number of atoms $N = \int |\phi|^2 dx$. This
equation is obtained in the case of a highly anisotropic external
potential under the assumption that the transversal trapping
potential is harmonic: $V(y,z) = m\omega_{\perp}^{2}(y^2 + z^2)/2
$ and $\omega_{\perp} \gg \omega_{x}.$  Under such conditions we
can consider the solution of 3D equation to have the form
$U(x,y,z;t) = R(y,z)\phi(x,t)$ where $R_{0}^2 =
m\omega_{\perp}\exp(-m\omega_{\perp} \rho^2/\hbar)/(\pi\hbar)$.
Averaging in the radial direction (i.e. integrating over the
transversal variables) we come to equation (\ref{gpe}) describing
the dynamics of the gas in longitudinal direction. The potential
$V(x,t)$ is assumed to be $V(x,t) = m\omega_{x}^{2}x^{2}F(t)$ ,
where $F(t)$ describes the time dependence of the potential, which
we consider here as $F(t)=1+h \sin(wt)$. The effective one
dimensional mean field nonlinearity coefficient $g_{1D} = 2\hbar
a_{s}\omega_{\perp}$, where $a_{s}$ is the atomic scattering
length. $a_{s} > 0$ corresponds to the Bose gas with a repulsive
interaction between atoms and $a_{s} < 0$ to an attractive
interaction.

We have the following dimensionless form of the equation
(\ref{gpe})
\begin{eqnarray}\label{gpe1}
i\psi_{t} + \frac{1}{2}\psi_{xx} -\frac{x^2}{2}F(t)\psi -
g|\psi|^{2}\psi +
\mu \psi = \nonumber\\
= i\gamma(-\frac{1}{2}\psi_{xx} + \frac{x^2}{2}F(t)\psi +
g|\psi|^{2}\psi -\mu \psi) = R(\psi,\psi^{\ast}),
\end{eqnarray}
by setting: $t = \omega_{x}t,$ $l = \sqrt{\hbar/(m\omega_{x})},$
$x= x/l,$ $\psi = \sqrt{2|a_{s}|\omega_{\perp}/\omega_{x}}\phi,$
with $g = \pm 1$ for the repulsive and attractive two-body
interactions respectively.

\section{Variational analysis}
To describe collective oscillations  of a Bose gas under damping
we employ the variational approach. For the wavefunction
$\psi(x,t)$ we use the gaussian trial function

\begin{equation}\label{Gaussian}
\psi(x,t)=A(t)exp(-\frac{x^2}{2a^2(t)}-\frac{ib(t)x^2}{2}-i\varphi(t)),
\end{equation}
where A, a, b and $\varphi$ are the amplitude, width, chirp and
linear phase, respectively.

Equation (\ref{gpe1}) can be obtained from the variational
equations
\begin{equation}\label{vareq}
\frac{\partial L}{\partial \psi^*} - \frac{\partial}{\partial x}
\frac{\partial L}{\partial \psi^*_x} - \frac{\partial}{\partial t}
\frac{\partial L}{\partial \psi^*_t} + \frac{\delta L_R}{\delta
\psi^*} = 0,
\end{equation}
where L is the Lagrangian density, $L \equiv L(x,t)  $, of a
conservative system, given by
\begin{equation}\label{Lagden}
L = \frac{i}{2}(\psi_t \psi^* - \psi^*_t \psi) -
\frac{1}{2}|\psi_x|^2 - (\frac{x^2}{2}F(t) - \mu)|\psi|^2 -
\frac{g}{2}|\psi|^4
\end{equation}
and $L_R$ is defined as $\partial L_R / \partial \psi^* =
-R(\psi,\psi^*)$, where $R(\psi,\psi^*)$ is the right side of
equation (\ref{gpe1}).

Inserting trial function(\ref{Gaussian}) into equation
(\ref{Lagden}) and averaging it as
\begin{equation}\label{averaging}
\bar{L} = \int{L(x,t)dx}
\end{equation}
we obtain the averaged Lagrangian of the conservative system in
terms of the trial function parameters:
\begin{eqnarray}\label{avLag}
\frac{\bar{L}}{\sqrt{\pi}}=\frac{A^2a^3b_t}{4} + A^2a\varphi_t -
\frac{A^2}{4a}- \frac{A^2a^3b^2}{4} -
\nonumber\\
- \frac{F A^2 a^3}{4} - \frac{g A^4 a}{2 \sqrt{2}} + \mu A^2 a.
\end{eqnarray}

Using equation (\ref{vareq}) and its conjugate, we obtain a system
of equations for the variational parameters $\eta_i$
\cite{And,FAGT}:
\begin{equation}\label{varpareq}
\frac{\partial \bar{L}}{\partial \eta_i} -
\frac{d}{dt}\frac{\partial \bar{L}}{\partial \eta_{it}} = \int dx
(R \frac{\partial \psi^*}{\partial \eta_i} + R^*\frac{\partial
\psi}{\partial \eta_i}).
\end{equation}

Inserting (\ref{Gaussian}) and (\ref{avLag}) into equation
(\ref{varpareq}) we derive the following system of ordinary
differential equations (ODE):
\begin{eqnarray}\label{ODE}
\frac{d(A^2 a)}{dt} = \frac{\gamma A^2}{2a} + \frac{\gamma A^2 a^3
b^2}{2} + \frac{\gamma F A^2 a^3}{2} +
\nonumber\\
+ \sqrt{2} \gamma g A^4 a - 2 \gamma \mu A^2 a,
\nonumber\\
\frac{d(A^2 a^3)}{dt} = -2 A^2 a^3 b - \frac{\gamma A^2 a}{2} +
\frac{3 \gamma A^2 a^5 b^2}{2} + \nonumber\\
+ \frac{3 \gamma F A^2 a^5}{2} + \frac{ \gamma g A^4
a^3}{\sqrt{2}} - 2 \gamma \mu A^2 a^3,
\nonumber\\
\frac{db}{dt} = \frac{2 \gamma b}{a^2} - \frac{1}{a^4} + b^2 + F -
\frac{g A^2}{\sqrt{2} a^2},
\nonumber\\
\frac{d \varphi}{dt} = - \frac{\gamma b}{2} + \frac{1}{2 a^2} +
\frac{5 g A^2}{4 \sqrt{2}} - \mu.
\end{eqnarray}

This system of equations can also be obtained from the modified
form of the conservation law for the number of atoms and by using
the moments method, as shown in Appendix. Let us rewrite the
system using the notations $x=a^2, y=A^2$:
\begin{eqnarray}\label{rewODE}
x_t = - 2xb - \gamma + \gamma x^2 b^2 + \gamma F x^2 -
\frac{\gamma g y x}{\sqrt{2}},
\nonumber\\
y_t = yb + \frac{\gamma y}{x} + \frac{5 \gamma g y^2}{2 \sqrt{2}}
- 2 \gamma \mu y,
\nonumber\\
b_t = \frac{2 \gamma b}{x} - \frac{1}{x^2} + b^2 +F -
\frac{gy}{\sqrt{2} x}.
\end{eqnarray}

This ODE system is the main result of this section.

\section{Numerical simulations}

We have carried out a series of time dependent simulations of the
system within the variational approach using equation
(\ref{rewODE}) and also by performing exact numerical calculations
using equation (\ref{gpe1}). In our numerical calculations we
discretize the problem in a standard way, with the time step $dt$,
and spatial step $dx$, so $\psi^k_j$ approximates $\psi(jdx,
kdt)$. More specifically we approximate the governing equation
(\ref{gpe1}) with the following semi-implicit Crank-Nickolson
scheme using split-step method \cite{Adh}
\begin{eqnarray}\label{splstep}
\frac{i(\psi^{k+1}_j-\psi^{k+\frac{2}{3}}_j)}{dt}+
\frac{i(\psi^{k+\frac{2}{3}}_j-\psi^{k+\frac{1}{3}}_j)}{dt}+
\nonumber\\
+\frac{i(\psi^{k+\frac{1}{3}}_j-\psi^k_j)}{dt}=
\frac{1}{2}H_1(\psi^{k+1}_j+\psi^{k+\frac{2}{3}}_j)+
\nonumber\\
+\frac{1}{2}H_2(\psi^{k+\frac{2}{3}}_j+\psi^{k+\frac{1}{3}}_j)+
\frac{1}{2}H_3(\psi^{k+\frac{1}{3}}_j+\psi^k_j),
\end{eqnarray}
where $H_1=0.5(1+i\gamma)(x^2 F(t)/2 + g|\psi(x,t)|^{2} -\mu),$
$H_2=-(1+i\gamma)\partial^2 / \partial x^2,$ $H_3=H_1$ and
$\psi^{k+\frac{2}{3}},$ $\psi^{k+\frac{1}{3}}$ are defined so that
\begin{eqnarray}\label{splstep1}
\frac{i(\psi^{k+\frac{1}{3}}_j-\psi^k_j)}{dt}=
\frac{1}{2}H_1(\psi^{k+\frac{1}{3}}_j+\psi^k_j),
\end{eqnarray}
\begin{eqnarray}\label{splstep2}
\frac{i(\psi^{k+\frac{2}{3}}_j-\psi^{k+\frac{1}{3}}_j)}{dt}=
\frac{1}{2}H_2(\psi^{k+\frac{2}{3}}_j+\psi^{k+\frac{1}{3}}_j),
\end{eqnarray}
\begin{eqnarray}\label{splstep3}
\frac{i(\psi^{k+1}_j-\psi^{k+\frac{2}{3}}_j)}{dt}=
\frac{1}{2}H_3(\psi^{k+1}_j+\psi^{k+\frac{2}{3}}_j).
\end{eqnarray}
Solving (\ref{splstep1}) and (\ref{splstep2}) first at time
$t_k=kdt$ we produce intermediate solutions $\psi^{k+\frac{1}{3}}$
and $\psi^{k+\frac{2}{3}}$. The final solution for one time step
dt is obtained from (\ref{splstep3}).

The results of numerical simulations of both PDE and ODE models
are presented below.


Figure \ref{equilstate} shows time evolution of the width and the
norm of the condensate for the case of $F(t)=1$, i.e. when the
external potential has no time perturbations. The figure is
presented to compare the character of the width oscillations and
the behavior of the norm depending on the values of the chemical
potential $\mu$ and the dissipative constant $\gamma$. As seen,
the frequency of the oscillations does not depend on the
dissipative constant at all and depends weakly on the chemical
potential, e.g. when $\mu=2$, $w_0=1.808$ while when $\mu=3$,
$w_0=1.77$. For greater values of $\gamma$, the oscillations damp
faster. We see that the ODE leads to an equilibrium state the norm
of which is 3-4 percent less than that of the PDE results.

From figure \ref{equilstate} it can be seen that damping process
of the condensate eventually leads to the equilibrium state. This
equilibrium state can also be obtained analytically by solving
equation (\ref{rewODE}). Taking into consideration that in an
equilibrium state $x_t=0$, $y_t=0$ and $b=0$ the following
expressions can be obtained:
\begin{eqnarray}\label{equilst}
a^2=\frac{4 \mu + \sqrt{16\mu^2+60}}{10},
\nonumber\\
A^2=\frac{4\sqrt{2}}{5}\mu - \frac{2\sqrt{2}}{5a^2}.
\end{eqnarray}
Putting the value of chemical potential e.g. for $\mu=2$ we find
$a=1.383$, $A=1.403$ and $N=A^2 a \sqrt{\pi} = 4.825$ which are
confirmed by the PDE results.


In figure \ref{norm} we can observe an interesting behavior of the
norm. If an external perturbation is applied to a trapped BEC
which is already in the equilibrium state then the norm of the
condensate starts decreasing. In the figure the frequency of the
periodical trap perturbation is taken to be equal to the
eigenfrequency of the system determined from the figure
\ref{equilstate} and the amplitude of time perturbations is taken
as $h=0.06$. We see that for smaller values of the dissipative
constant $\gamma$ the norm goes farther from the equilibrium
state. That is, the dissipative constant is not simply the
quantity which is responsible for diminishing the norm, but it is
the constant keeping a condensate in the equilibrium state.

The width dynamics under main resonance with the initial wave
packet taken in the equilibrium state is depicted in figure
\ref{widthres}. As shown, in contrast to the norm the width
oscillates near the previous point. As seen from the figure the
amplitude of the width oscillations becomes stable by the time
$t=140$.


Performing ODE and PDE calculations for the frequencies which lie
around the eigenfrequency of the BEC and measuring the amplitudes
of the above-said stable oscillations we have plotted the values
of the oscillation amplitude as a function of the frequency of the
periodical trap perturbations with $h=0.03$ and $h=0.06$ in the
cases  $\gamma=0.01$ and $\gamma=0.005$ in figure \ref{ampfrchar}.


We see that e.g. when $h=0.06$ and $\gamma=0.005$ the highest
amplitude of oscillations is driven in the trap perturbation with
the frequency $w=1.89$ which is more than the eigenfrequency
$w_0=1.806$. Bistability appears with the smaller values of
$\gamma$ in the vicinity of this critical frequency. For $w=1.88$
we observe large oscillations with $a_{osc}=1.6$, while at
$w=1.91$ we observe much smaller width oscillations with
$a_{osc}=0.6$. It can be seen that with the growth of trap
perturbations the value of the critical frequency becomes greater.

Let us estimate parameters for the experiment. The magnetic trap
can be taken with parameters $\omega_\perp = 2 \pi \times 40$ Hz,
$\omega_x = 2 \pi $ Hz, and the number of atoms of $^{85}$Rb
$N=0.23 \times 10^4$. For the external field $B=161.57$, $a_s=0.3
nm$ (repulsive gas). Then, the eigenfrequency of the harmonically
trapped BEC $2 \pi \nu_0 = 1.808 \times \omega _x = 11,36 $ Hz. By
applying the external perturbation $F(t)=1+0.06 \sin(2 \pi \nu t)$
with $2 \pi \nu=11.36 $ Hz to the trapped BEC one will observe
decreasing of the number of atoms by 20, 13 and 7 percent for
$\gamma$ = 0.1, 0.2 and 0.3 respectively. For the external
perturbation with $2 \pi \nu = 11.8 $ Hz, large oscillations will
be observed with $a_{osc} = 1.6l = 1.6 \sqrt{ \hbar /(m
\omega_x)}=17.4 \mu m$, while at $2 \pi \nu = 12 $ Hz one will
observe much smaller oscillations with $a_{osc} = 0.6 l = 6.5 \mu
m$.

\section{Conclusions}

In this paper we study the collective oscillations of a quasi one
dimensional Bose gas in the presence of dissipative effects. The
modified Gross-Pitaevskii equation in the framework of  the
phenomenological approach \cite{LPP,Burnett} has been used. To
describe evolution of oscillations we employ the modified
variational approach taking into account the dissipation. We
confirm results for the oscillations damping obtained from the
system of equations for the wave function parameters, by the
moments method and direct numerical simulations of the full GP
equation.

The calculations show that damping oscillations of a BEC which
eventually comes to an equilibrium state occur with the
eigenfrequency which does not depend on the value of the
dissipative constant $\gamma$ and only depends on the chemical
potential $\mu$.

The expressions for the width and the norm of a condensate in an
equilibrium state have been derived analytically.

We study the main resonance in condensate oscillations. For a BEC
in an equilibrium state it is shown that making the external
potential oscillate leads to decreasing of the condensate norm and
the norm begins oscillating around the point which is less than
the previous stationary one in the equilibrium state, whereas the
condensate width oscillates near the previous point.

We have also shown that in resonances the bistability appears with
smaller values of the dissipative constant in the vicinity of the
critical frequency which is above the eigenfrequency of the BEC.

\ack

The work was partially supported by a fund for fundamental
research support from the Uzbek Academy of sciences (Award N
17-04). FKhA is grateful to the Physics Department of the
University of Salerno for the research grant.

\appendix
\section*{Appendix}
\setcounter{section}{1}

Differentiating the norm of the condensate by time we can obtain
the first equation of the system (\ref{rewODE}) as
\begin{equation}
\label{app1} \frac{dN}{dt} = \frac{d}{dt} \int |\psi|^2 dx =
\int(\psi^*_t \psi + \psi^* \psi_t) dx.
\end{equation}

Inserting $\psi_t$ from equation (\ref{gpe1}) into (\ref{app1}) we
derive the modified form of the conservation law for the norm of
the condensate [29].
\begin{eqnarray}
\label{app2}
\frac{dN}{dt} = \gamma \int|\psi_x|^2 dx + \gamma F
\int x^2 |\psi|^2 dx +
\nonumber\\
+ 2 \gamma g \int|\psi|^4 dx - 2 \gamma \mu \int|\psi|^2 dx.
\end{eqnarray}
Substituting the Gaussian trial function (\ref{Gaussian}) into
this equation we obtain equation (\ref{rewODE}a).

Equations (\ref{rewODE}b) and (\ref{rewODE}c) can be obtained by
the moments method. We obtain equation (\ref{rewODE}b) by
calculating
\begin{equation}
\label{app3} \frac{d<x^2>}{dt} = \int(\psi^*_t x^2 \psi + \psi^*
x^2 \psi_t) dx.
\end{equation}
The substitution of the $\psi_t$ from equation (\ref{gpe1}) leads
to the following equation:
\begin{eqnarray}
\label{app4}
\frac{d<x^2>}{dt} = i \int x (\psi^*_x \psi - \psi_x
\psi^*) dx + \gamma \int x^2 |\psi_x|^2 dx -
\nonumber\\
- \gamma \int|\psi|^2 dx + \gamma F \int x^4 |\psi|^2 dx +
\nonumber\\
+ 2 \gamma g \int x^2 |\psi|^4 dx - 2 \gamma \mu \int x^2 |\psi|^2
dx.
\end{eqnarray}
Substituting trial function (\ref{Gaussian}) we get equation
(\ref{rewODE}b). Equation (\ref{rewODE}c) is derived by
calculating
\begin{eqnarray}
\label{app5}
\frac{d<p^2>}{dt} = - \frac{d}{dt} \int \psi^*
\psi_{xx} dx =
\nonumber\\
= - \int(\psi^*_t \psi_{xx} + \psi^* \psi_{txx}) dx.
\end{eqnarray}
Integrating by parts and substituting $\psi_t$ from equation
(\ref{gpe1}) we obtain
\begin{eqnarray}
\label{app6}
\frac{d}{dt} \int |\psi_x|^2 dx = i F \int x (\psi^*
\psi_x - \psi \psi^*_x) dx +
\nonumber\\
+ ig \int (\psi^{*2} \psi_x^2 - \psi^2 \psi^{*2}_x) dx +
\nonumber\\
+ \gamma \int |\psi_{xx}|^2 dx - \gamma F \int |\psi|^2 dx +
\nonumber\\
+ \gamma F \int x^2 |\psi_x|^2 dx + 4 \gamma g \int |\psi|^2
|\psi_x|^2 dx +
\nonumber\\
+ \gamma g \int (\psi^{*2} \psi_x^2 + \psi^2 \psi^{*2}_x) dx - 2
\gamma \mu \int |\psi_x|^2 dx.
\end{eqnarray}

Inserting the Gaussian ansatz into this equation and using results
(\ref{rewODE}a) and (\ref{rewODE}b) we obtain equation
(\ref{rewODE}c).

\Figures

\begin{figure}[h]
\caption{ Dynamics of the norm (a) and the width (b) of the
repulsive BEC in a harmonic trap without perturbation. The upper
and lower diagrams are plotted for the cases $\mu=3$ and $\mu=2$
respectively. Solid and dotted lines show PDE and ODE results.}
\label{equilstate}
\end{figure}

\begin{figure}[h]
\caption{The behavior of the norm of the trapped BEC when the trap
starts oscillating with the amplitude $h=0.06$. Before the trap
oscillations BEC was in the equilibrium state. We compare the PDE
and ODE system simulations. The solid lines stand for full
numerical simulations of the PDE, while the dotted lines represent
the ODE results.}
\label{norm}
\end{figure}

\begin{figure}[h]
\caption{BEC width versus time t in the main resonance in the case
$h=0.06$, $\mu=2$, $\gamma=0.01$. Solid and dotted lines show the
PDE and ODE results respectively.} \label{widthres}
\end{figure}

\begin{figure}[h]
\caption{Oscillation amplitude as predicted by the PDE (scatter)
and ODE (line) models. Here $\gamma=0.005$ (solid lines and
squares) and $\gamma=0.01$ (dotted lines and circles). The upper
two lines are for the case $h=0.06$, the lower two are for the
case $h=0.03$. The chemical potential is equal to 2.}
\label{ampfrchar}
\end{figure}

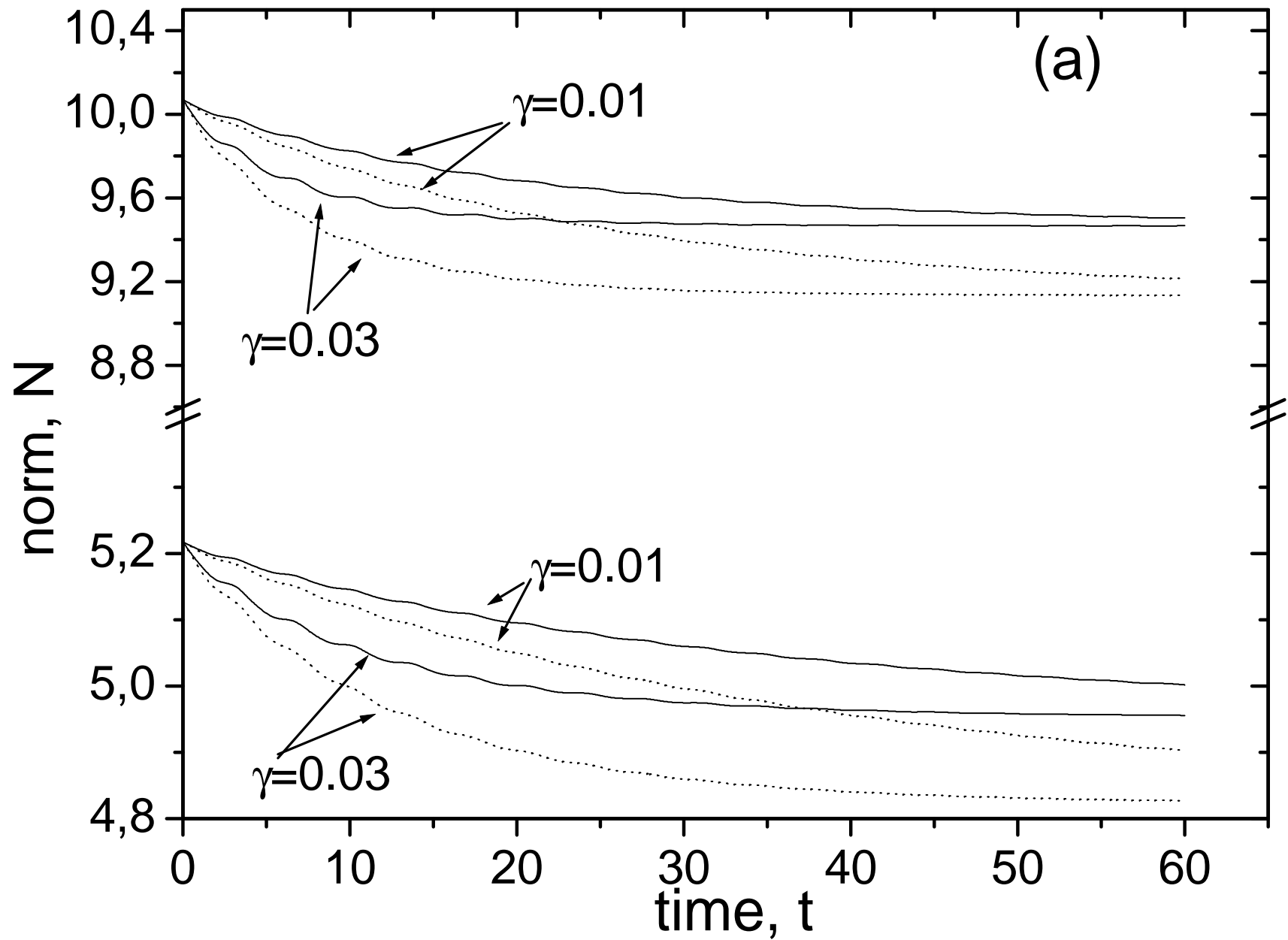
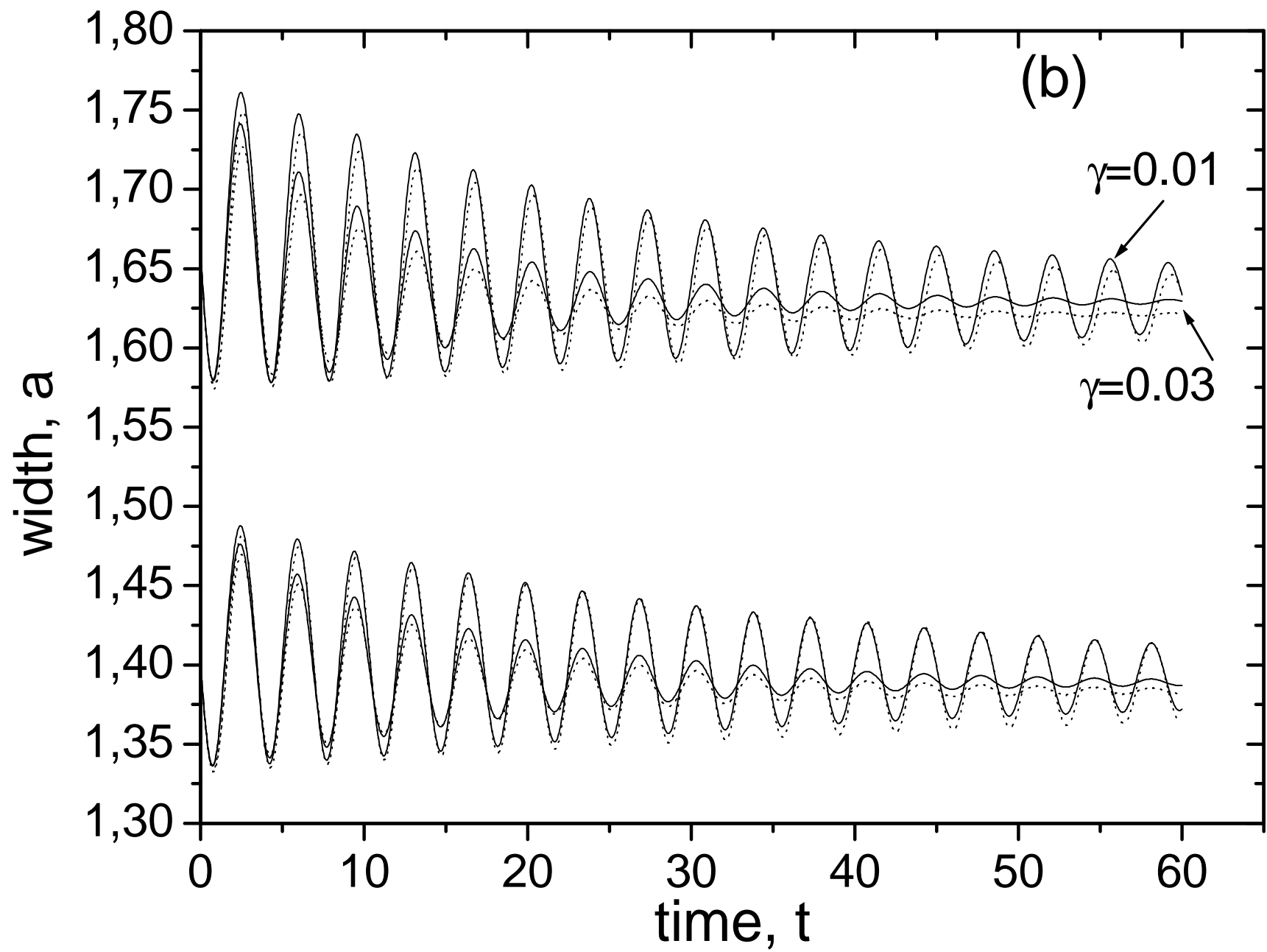
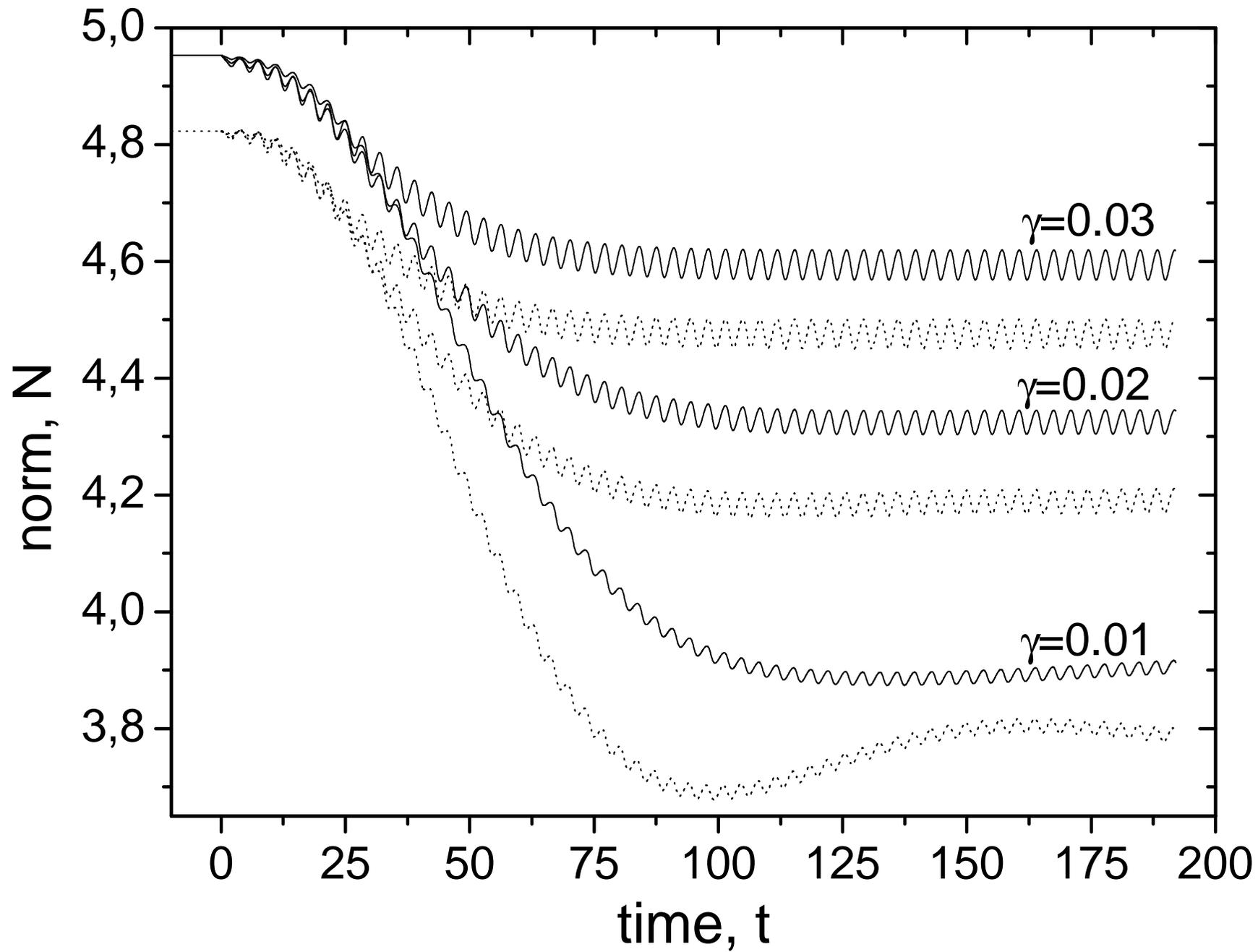
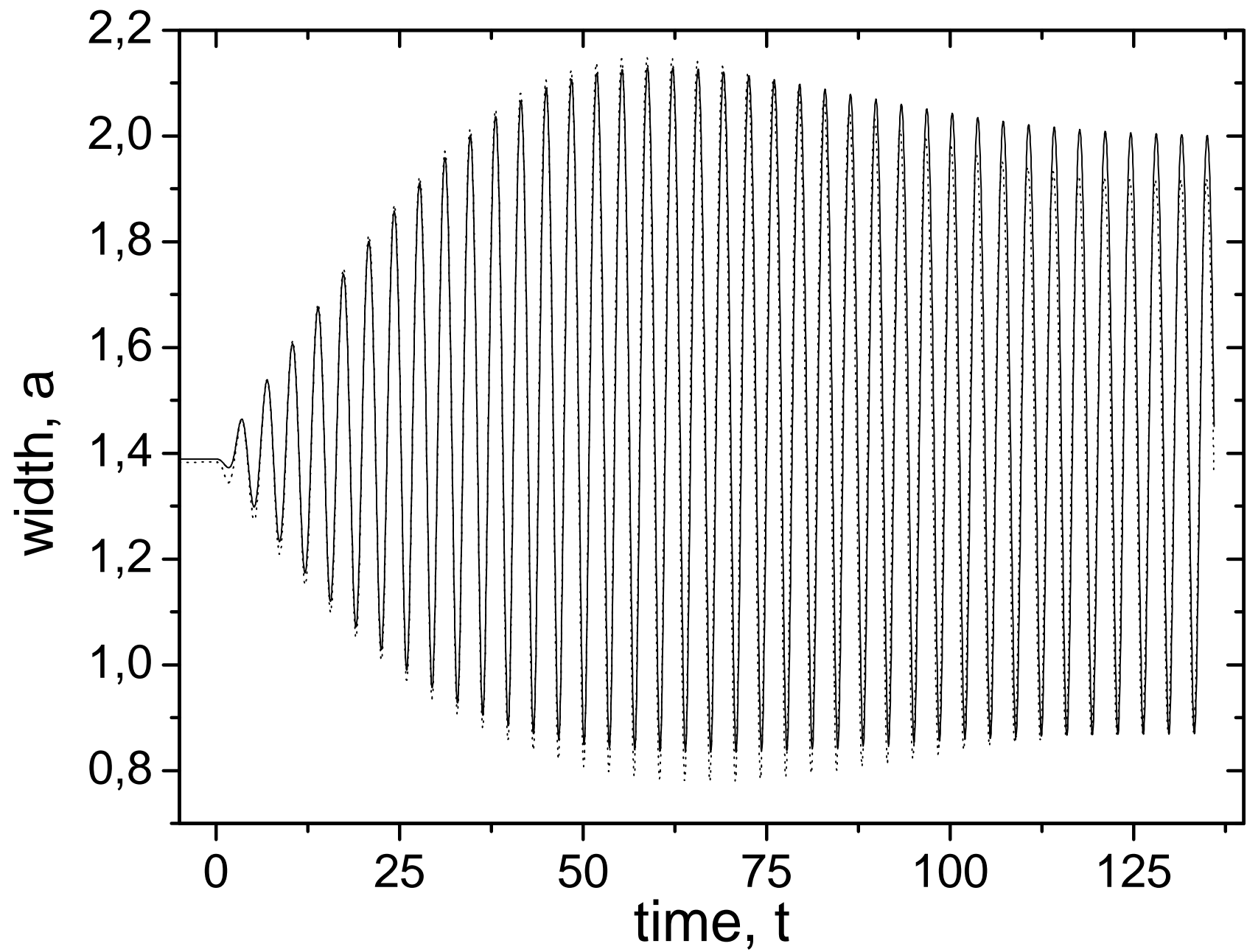
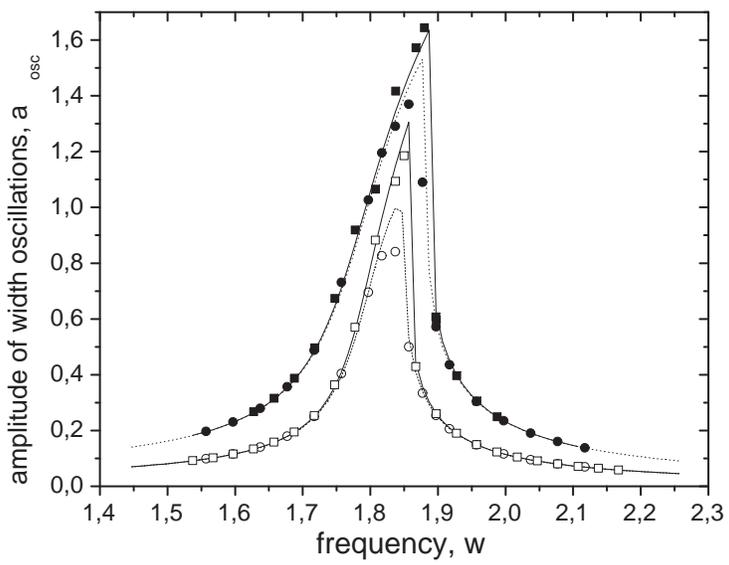
\end{document}